\documentclass[prl, twocolumn, superscriptaddress]{revtex4-1}
\usepackage{amsmath}
\usepackage[english]{babel}
\usepackage[pdftex]{graphicx}
\usepackage{color}

\usepackage[sort&compress]{natbib}
\usepackage{hyperref}
\graphicspath{{images/}{./}}

\usepackage{amsmath}
\usepackage{amssymb}
\usepackage{mathtools}
\usepackage{braket}

\renewcommand{\vec}[1]{\ensuremath{\boldsymbol{#1}}}

\renewcommand{\deg}{^{\circ}}

\begin{document}

\title{Generalized Hamiltonian for Kekul\'e graphene \\
and the emergence of valley-cooperative Klein tunneling}

\author{Santiago Galv\'an y Garc\'ia}
\email{santiagogyg@icf.unam.mx}
\affiliation{Instituto de Ciencias F\'isicas, Universidad Nacional Aut\'onoma de M\'exico, Cuernavaca 62210, M\'exico}

\author{Thomas Stegmann}
\email{stegmann@icf.unam.mx}
\affiliation{Instituto de Ciencias F\'isicas, Universidad Nacional Aut\'onoma de M\'exico, Cuernavaca 62210, M\'exico}

\author{Yonatan Betancur-Ocampo}
\email{ybetancur@icf.unam.mx}
\affiliation{Instituto de Ciencias F\'isicas, Universidad Nacional Aut\'onoma de M\'exico, Cuernavaca 62210, M\'exico}

\begin{abstract}
We introduce a generalized Hamiltonian describing not only all topological phases observed experimentally in Kekul\'e graphene (KekGr) but predicting also new ones. These phases show features like a quadratic band crossing point, valley splitting, or the crossing of conduction bands, typically induced by Rashba spin-orbit interactions or Zeeman fields. The electrons in KekGr behave as Dirac fermions and follow pseudo-relativistic dispersion relations with Fermi velocities, rest masses, and valley-dependent self-gating. Transitions between the topological phases can be induced by tuning these parameters. The model is applied to study the current flow in KekGr $pn$ junctions evidencing a novel cooperative transport phenomenon, where Klein tunneling goes along with a valley flip. These junctions act as perfect filters and polarizers of massive Dirac fermions, which are the essential devices for valleytronics. The plethora of different topological phases in KekGr may also help to establish phenomena from spintronics.
\end{abstract}

\maketitle

Over the last two decades, graphene has attracted the scientific community due to its outstanding properties \cite{Novoselov2005}. One of them is the fact that the electrons in graphene present, apart from the intrinsic spin, two pseudospins: the sublattice and the valley pseudospin \cite{Neto2009, Bercioux2015}. The latter one may be used for a new type of electronics called valleytronics \cite{Rycerz2007, Schaibley2016, Gunlycke2011, Hossain2021}. In order to fabricate valleytronic devices, it has been shown that the valley pseudospin can be manipulated by strain and defects \cite{Rycerz2007, Gunlycke2011, Stegmann2016, Stegmann2018, Low2010, Zhao2020, Fujita2010, GarciaPomar2008, Zhong2017}. Another strategy for valleytronics is to break the chiral symmetry in graphene through a Kekul\'e distortion \cite{Gutierrez2016, Liu2017, Eom2020, Giovannetti2015, Bao2021}. This distortion consists of alternating hopping parameters and on-site energies, leading to an enlarged unit cell of six atoms \cite{Chamon2000, Konschuh2010, Hou2007, Liu2017, Wu2016, Eom2020, Pachoud2014, Giovannetti2015, Ren2015, Venderbos2016, Andrade2019, Wang2018, Wang2020, Gamayun2018, Mojarro2020}. Currently, two topological phases in Kekul\'e graphene (KekGr) have been predicted and confirmed experimentally \cite{Bao2021, Gutierrez2016, Virojanadara2010, Eom2020}. Moreover, KekGr may establish a bridge between valleytronics and spintronics to realize intriguing phenomena predicted in semiconductors \cite{Datta1990, Mucciolo2002, Khodas2004, Bercioux2010, Marchenko2012, Asmar2013, Han2014, Bercioux2015, Avsar2020}. Note that the chiral symmetry breakig observed in KekGr is also of crucial importance to emulate mass generation of elementary particles in condensed matter, as studied in the standard model \cite{Nambu1961, Obertelli2021}.

Recently, Kekul\'e distortions were generated by intercalating Li atoms between a graphene sheet and a SiC substrate \cite{Bao2021, Virojanadara2010}, confirming the chiral symmetry breaking predicted by effective models based on a tight-binding approach \cite{Gamayun2018, Beenakker2018}. Nevertheless, the current models do not explain some topological phases recently discovered in experimental studies of KekGr\cite{Bao2021}. Here, we propose a generalized Hamiltonian that embodies all the known results in KekGr and predicts various new topological phases. This generalized Hamiltonian is used to study the transport properties of KekGr $pn$ junctions. We observe a valley-flip process due to the valley-dependent Fermi velocity, giving rise to a novel Klein tunneling. This effect is persistent against variations of the Fermi level and electrostatic gating, and emerges in all topological phases of KekGr within the ultra-pseudo-relativistic limit. Moreover, the proposed devices fully valley polarize the current flow (in certain parameter regime) and hence, are of relevance for valleytronic applications.

\begin{figure*}[t]
 \includegraphics[scale=0.47]{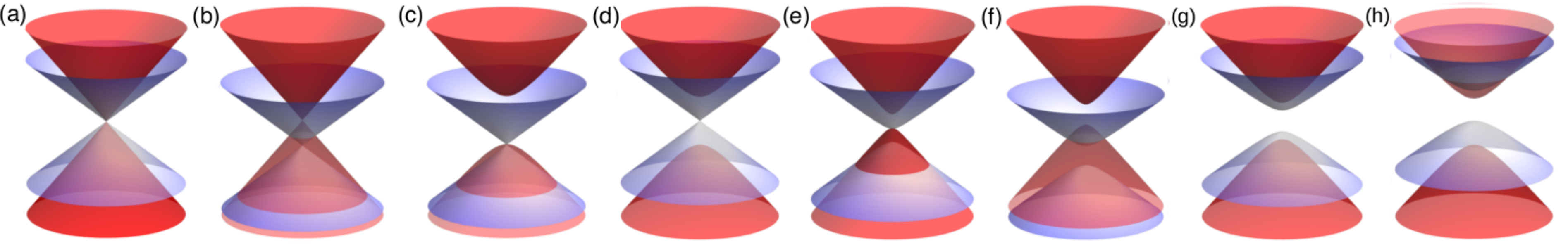}
\caption{Topological phases of Kekul\'e graphene (KekGr). (a) and (b) Massless Dirac fermions in both valleys. (c) and (d) Chiral symmetry breaking in a single valley, where electrons behave as massless and massive Dirac fermions. (e) and (f) Quadratic band crossing point and valley-orbit coupling, respectively. (g) Zeeman-like effect. (h) Crossing of conduction bands.}
\label{DCs}
\end{figure*}

\begin{figure}
    \centering
    \includegraphics[scale=0.5]{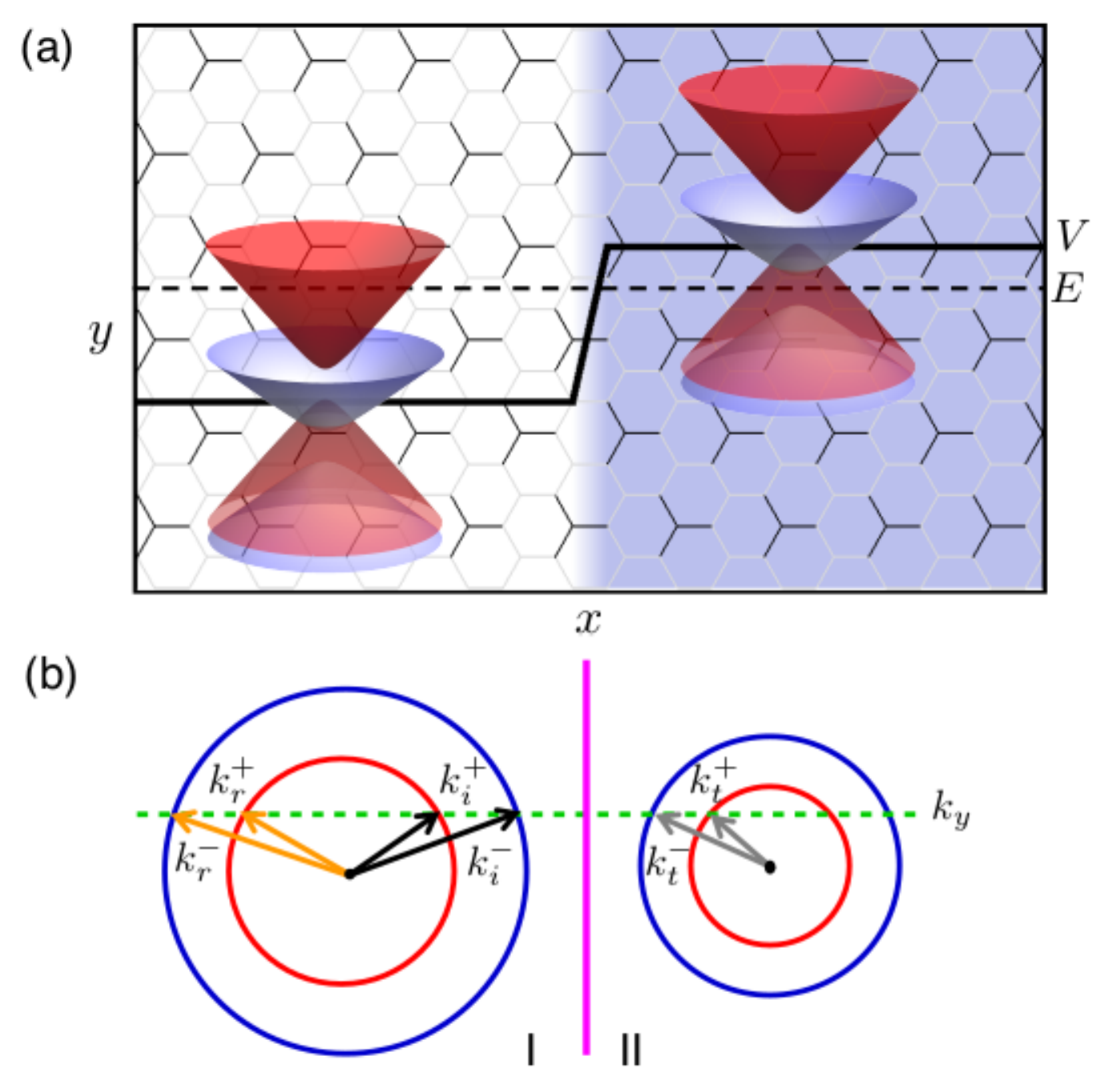}
    \caption{KekGr $pn$ junction. (a) Potential profile of the $pn$ junction showing the shift of the bandstructure in the two regions. (b) Kinematical construction that shows the conservation of energy $E$, momentum $k_y$, and current density. The arrows represent the wave vectors $\vec{k}^\nu$ for the states involved in the scattering process.}
    \label{fig:junction}
\end{figure}

KekGr shows bond distortions of Y and O shape, known as Kek-Y and Kek-O textures. This rearrangement in the lattice causes drastic changes in the electronic band structure. The Dirac cones in the valleys K$^+$ and K$^-$ are folded towards the $\Gamma$ point in the Brillouin zone \cite{Giovannetti2015, Gamayun2018, Andrade2019, Konschuh2010, Venderbos2016}. In the case of the Kek-O texture, the low-energy regime shows degenerated energy bands separated by a gap, while in the case of the Kek-Y, the low-energy excitations follow a non-degenerated and valley-dependent linear dispersion relation \cite{Gutierrez2016}. The tuning of the onsite energies of specific atoms in the unit cell breaks the chiral symmetry for obtaining simultaneously massless and massive Dirac fermions. Here, we propose the most general effective Hamiltonian of KekGr
\begin{eqnarray}
    H_\textrm{KekGr} & = & \left(\begin{matrix}
    2\lambda & v_\sigma p_- & v_\tau p_- & 2\eta\frac{p_-}{p_+}\\
     v_\sigma p_+ & 2\rho & 2\delta & v_\tau p_-\\
     v_\tau p_+ & 2\delta & 2\rho & v_\sigma p_-\\
     2\eta\frac{p_+}{p_-} & v_\tau p_+ & v_\sigma p_+ & 2\lambda 
    \end{matrix}\right)
    \label{HKGS}
\end{eqnarray}

\noindent to describe the superposition of two different massive Dirac fermions in the two valleys. In the Hamiltonian \eqref{HKGS}, $v_\sigma$ is the Fermi velocity of pristine graphene, $v_\tau$ is the velocity that quantifies the magnitude of the Kekul\'e distortion, and $\vec{p} = (p_x,p_y)$ is the linear momentum with $p_\pm = p_x \pm ip_y$.  The parameters $\lambda$, $\rho$, and $\delta$ are related with the onsite potential and break the chiral symmetry for one or both valleys, while the parameter $\eta$ quantifies the spin-orbit interaction. The eigenenergies of the Hamiltonian \eqref{HKGS} are
\begin{equation}\label{Esnu}
E^\nu_s = \xi_\nu + s\sqrt{(v_\sigma + \nu v_\tau)^2p^2 + \mu^2_\nu},   
\end{equation}

\noindent where $s = \textrm{sign}(E - \xi_\nu)$ and $\nu = \pm 1$ are the band and valley index, respectively. The self-gating and rest energies $\xi_\nu$ and $\mu_\nu$ depend on the parameters of the Hamiltonian \eqref{HKGS} as $\xi_\nu = \lambda + \rho + \nu(\delta + \eta)$ and $\mu_\nu = -\delta + \eta + \nu(\lambda -\rho)$, respectively. Tuning these parameters leads to different phases, which can be obtained in KekGr modulating the onsite energies, hopping parameters, and spin-orbit coupling, as shown in tight-binding calculations \cite{Giovannetti2015}. In general, electrons in KekGr behave like a mixture of massive Dirac fermions with relatively shifted dispersion relations identical to relativistic particles having limit velocities $v_\sigma + \nu v_\tau$. In Fig. \ref{DCs}, we show a gallery of topological phases such as massless Dirac fermions in both valleys, chiral symmetry breaking in a single valley, quadratic band crossing point, field-free Zeeman-like effect, and valley-orbit coupling. Atypical phases are obtained, for instance, the crossing of the conduction bands of the two valleys. 
 
To study the electronic transport of KekGr in the multiple possible phases displayed in Fig. \ref{DCs}, we make a plane wave ansatz using the eigenstates of the Hamiltonian \eqref{HKGS}
\begin{equation}
\vec{\Psi}^\nu_s(\vec{k},\vec{r}) = \vec{u}^\nu_s(\vec{k}) \textrm{e}^{i\vec{k}\cdot\vec{r}}
\end{equation}
where
\begin{equation}
\vec{u}^\nu_s(\vec{k})=  \frac{\left(1,\alpha^\nu_s\textrm{e}^{i\phi(\vec{k})},\nu\alpha^\nu_s\textrm{e}^{i\phi(\vec{k})},\nu\textrm{e}^{2i\phi(\vec{k})}\right)}{\sqrt{2[1 + (\alpha^\nu_s)^2]}},
\end{equation}
\begin{equation}
\alpha^\nu_s=  \frac{E^\nu_s -\xi_\nu-\nu\mu_\nu}{\sqrt{(E^\nu_s-\xi_\nu)^2-\mu^2_\nu}},\quad
\phi(\vec{k})= \arctan\left(\frac{k_y}{k_x}\right).
\end{equation}

\noindent The linear momentum $\vec{p}$ has the eigenvalue $p = \hbar k$ and $\vec{u}^\nu_s(\vec{k})$ are the four spinors of the Hamiltonian \eqref{HKGS}. The pseudospin angle is $\phi(\vec{k})$ and $\alpha^\nu_s$ is the generalization of the band index $s$ of massless Dirac fermions. 

We consider a KekGr $pn$ junction where the electrons, injected at energy $E$, impinge an electrostatic potential $V$, see Fig.~\ref{fig:junction}(a). The electrons are injected as a superposition of states from both valleys 
\begin{eqnarray}\label{wi}
    \vec{w}_s(\vec{r}) & = & \sum_{\nu} a^\nu\vec{u}^\nu_s(\vec{k}^\nu_i)\textrm{e}^{i\vec{k}^\nu_{i}\cdot\vec{r}},
\end{eqnarray}
\noindent where $\vec{k}^\nu_i = (k^\nu_{i,x},k_y)$ is the incident wave vector. The component $k_y$ is the conserved linear momentum along the interface, which is parametrized in terms of the incidence angle $\theta$ of electrons in the valley $K^+$
\begin{equation}
k_y = \frac{\sqrt{(E - \xi_+)^2 - \mu^2_+}}{\hbar(v_\sigma+v_\tau)}\sin\theta. 
\end{equation}

\begin{figure*}[t!!]
\includegraphics[scale=0.17]{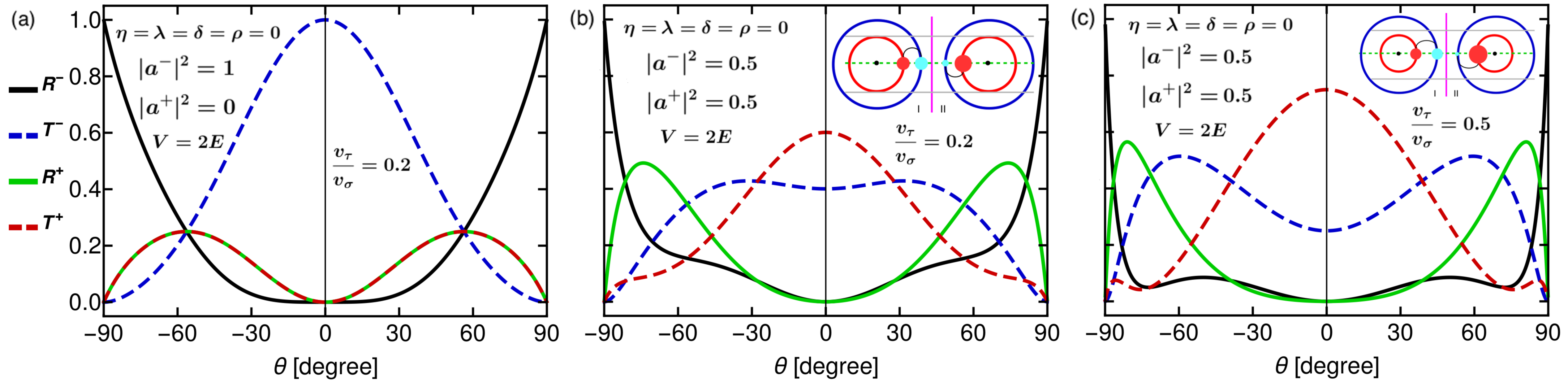}
\caption{(a)-(c) Reflection and transmission probabilities as a function of the incidence angle for different Kekul\'e distortions and valley polarizations of the injected electrons. The insets in (b) and (c) show the kinematical construction to illustrate the valley-cooperative Klein tunneling. The incident and transmitted electrons are represented by the superposition of electronic states in the valleys $K^\nu$, where the size of the circles indicates the occupation of the valleys.}
\label{FVCKT}
\end{figure*}

\noindent The amplitudes $a^\nu$ indicate the probability weight for the valley $K^\nu$. For instance, $a^+ = 1$ and $a^- = 0$ correspond to a full polarization of the injected electrons in the valley $K^+$. These electrons impinge the interface with different incidence angles because the component $k_x$ depends on the valleys, as shown in Fig. \ref{fig:junction}(b). Therefore, the wave function in region I is
\begin{eqnarray}
    \vec{\Psi}^\textrm{I}(\vec{r}) & = &  \vec{w}_s(\vec{r}) + \sum_\nu\vec{u}^\nu_s(\vec{k}^\nu_r) r^\nu\textrm{e}^{i\vec{k}^\nu_{r}\cdot\vec{r}},
\end{eqnarray}

\noindent where $\vec{k}^\nu_r = (k^\nu_{r,x},k_y)$ are the wave vectors of the reflected electrons and $r^\nu$ are the reflected amplitudes. In region II of the $pn$ junction, where the electrostatic potential shifts the dispersion relation, we have the wavefunction 
\begin{equation}
    \vec{\Psi}^\textrm{II}(\vec{r}) = \sum_\nu t^\nu\vec{u}^\nu_{s'}(\vec{k}^\nu_t)\textrm{e}^{i\vec{k}^{\nu}_{t}\cdot\vec{r}},
\end{equation}

\noindent with the band index $s' = \textrm{sign}(E - \xi_\nu - V)$. The transmitted wave vectors and amplitudes are $\vec{k}^\nu_t = (k^\nu_{t,x},k_y)$ and $t^\nu$, respectively. The continuity of the wave function $\vec{\Psi}^\textrm{I}(\vec{k},\vec{r})$ and $\vec{\Psi}^\textrm{II}(\vec{k},\vec{r})$ at the interface of the $pn$ junction allows determining the reflection and transmission coefficients
\begin{eqnarray}
R^\nu & = & -\frac{J^\nu_{s,x}(\vec{k}^\nu_r)}{\sum_\nu|a^\nu|^2J_{s,x}(\vec{k}^\nu_i)}|r^\nu|^2,\\
T^\nu & = & \frac{J^\nu_{s',x}(\vec{k}^\nu_t)}{\sum_\nu|a^\nu|^2J_{s,x}(\vec{k}^\nu_i)}|t^\nu|^2,\label{Tnu}
\end{eqnarray}
with the probability current density 
\begin{equation}
J^\nu_{s,x}(\vec{k}) = \frac{\partial E^\nu_s}{\partial p_x} = \frac{2\alpha^\nu_s(v_\sigma + \nu v_\tau)}{1 + (\alpha^\nu_s)^2}\cos{\phi(\vec{k})}.
\end{equation}
The reflection and transmission coefficients satisfy $\sum_\nu (R^\nu + T^\nu) = 1$ due to the conservation of $J_x$.

\begin{figure*}[t!!]
\includegraphics[scale=0.17]{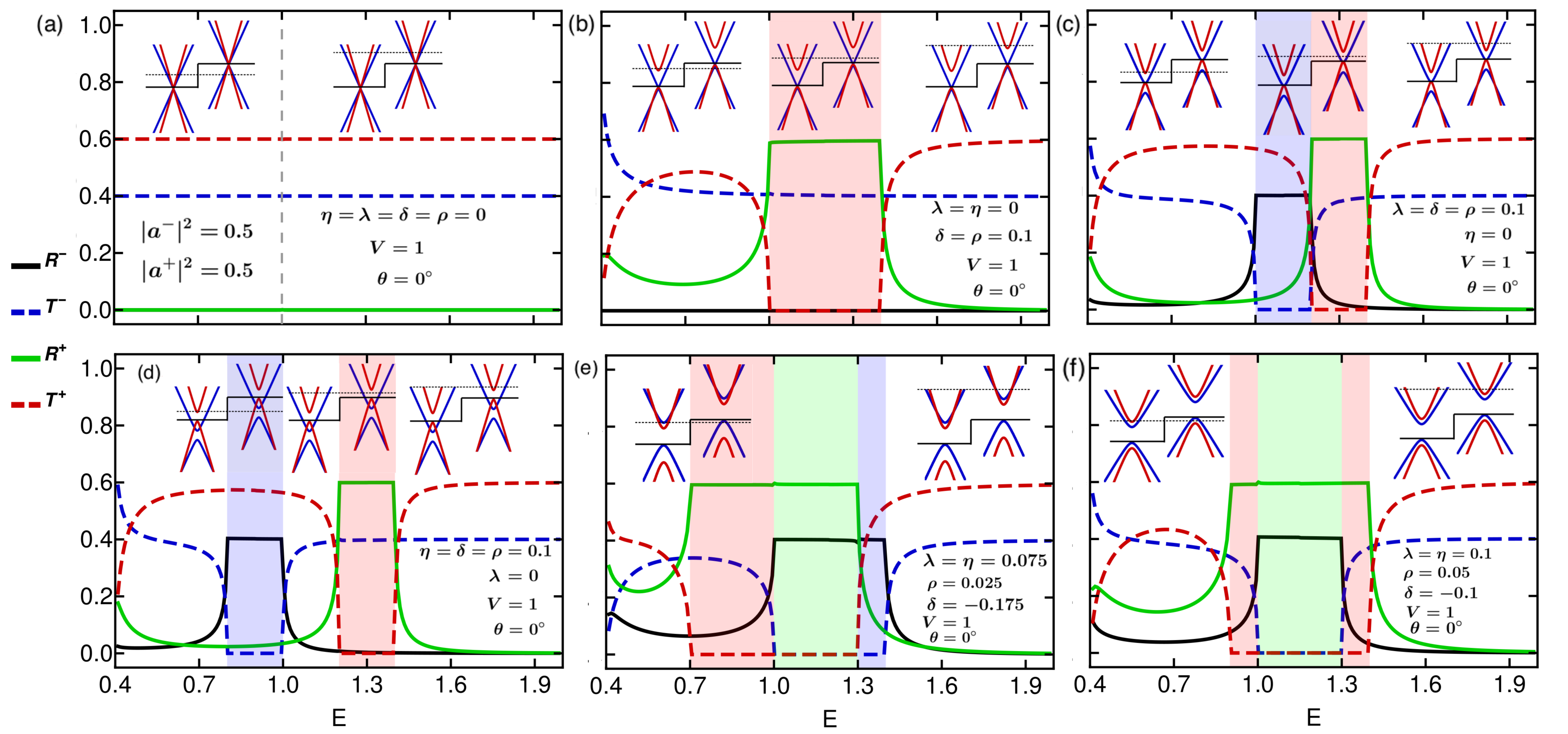}
\caption{Reflection and transmission probabilities as a function of the Fermi energy $E$ under normal incidence for different KekGr $pn$ junctions. (a) Massless Dirac fermions in both valleys, the dashed gray line separates the interband ($E < 1$) and intraband ($E > 1$) tunneling regime. (b) Chiral symmetry breaking in a single valley, (c) quadratic band crossing point, (d) valley-orbit coupling, (e) crossing conduction bands, and (f) Zeeman-like effect. The insets show the band structure of the junction, indicating $E$ by a dotted line. The red and blue rectangles represent the regions of perfect valley filtering. The green regions indicate the transport gap.}
\label{PVFE}
\end{figure*}

We analyze the case of massless Dirac fermions with $\mu_\nu = 0$ in the valleys $K^-$ and $K^+$. Figure~\ref{FVCKT}(a) shows the behavior for the reflection and transmission coefficients when the incident state is fully polarized in the valley $K^-$. For normal incidence, perfect tunneling appears due to the conservation of the pseudospin, while electrons with incidence angles $\theta \neq 0\deg$ have non-zero probabilities for the reflection and transmission in both valleys. The situation is identical if we inject electrons polarized in the valley $K^+$. A counter-intuitive transmission phenomenon occurs when the normally incident electrons occupy both valleys, $|a^\nu|^2 = 0.5$. In this case back-scattering is absent ($R^\nu=0$) but contrary to the expectation of finding an transmitted current that is distributed equally in both valleys, we obtain electron currents in the fractions $0.6$ and $0.4$ for the valleys $K^-$ and $K^+$, respectively, as shown in Fig. \ref{FVCKT}(b). $10 \%$ of the electrons in valley $K^-$ make a valley-flip to cross {\it cooperatively} the interface without backscattering. We name this unusual and novel transport phenomenon {\it valley-cooperative Klein tunneling}, as sketched in the insets of Fig.~\ref{FVCKT}(b).

The valley flip process is controlled only by the Kekul\'e distortion. Increasing the velocity $v_\tau$, the difference of $T_+ - T_-$ under normal incidence also increases, as shown in Fig.~\ref{FVCKT}(c). This novel effect is due to the Kekul\'e distortion and the conservation of the pseudospin. It is independent of the Fermi level and the electrostatic potential, see Fig.~\ref{PVFE}(a). To understand the valley-cooperative Klein tunneling, we calculate the transmission probabilities for normal incidence and weights $a^\pm$
\begin{equation}
T^\pm(0) = \frac{(v_\sigma \pm v_\tau)|a^\pm|^2}{\sum_\nu(v_\sigma + \nu v_\tau)|a^\nu|^2},
\label{VCKT}
\end{equation}

\noindent where the pseudospin angles are $\phi^-_i = \phi^+_i = 0$, $\phi^-_r = \phi^+_r = \pi$, and $\phi^-_t = \phi^+_t = (s-s')\pi/2$. For the reflection coefficients, we have $R^{\pm} = 0$. For $|a^\nu|^2 = 0.5$ the Eq.~\eqref{VCKT} can be simplified to $T^\pm(0) = \tfrac{1}{2}(1 \pm \tfrac{v_\tau}{v_\sigma})$. Clearly, we observe that $T^-(0) + T^+(0) = 1$. We recover the conventional Klein tunneling in graphene making $v_\tau = 0$, and as expected, unpolarized current under normal incidence crosses perfectly the electrostatic potential \cite{Katsnelson2006, Allain2011}. The extreme case $v_\sigma = v_\tau$ indicates that all the electrons in the valley $K^-$ perform a valley-flip to cross perfectly the junction, resulting in a fully valley polarized current. Moreover, $v_\sigma = v_\tau$ merges the conduction and valence bands in the valley $K^-$ to a flat band, giving rise to super-Klein tunneling, which is a typical phenomenon in the scattering of pseudospin-one particles \cite{Shen2010, Urban2011, Xu2016, BetancurOcampo2017, Nandy2019, Kim2020, ContrerasAstorga2020, Xu2021, Wang2021, Mandhour2020}. 

\begin{figure*}[t!!]
\includegraphics[scale=0.17]{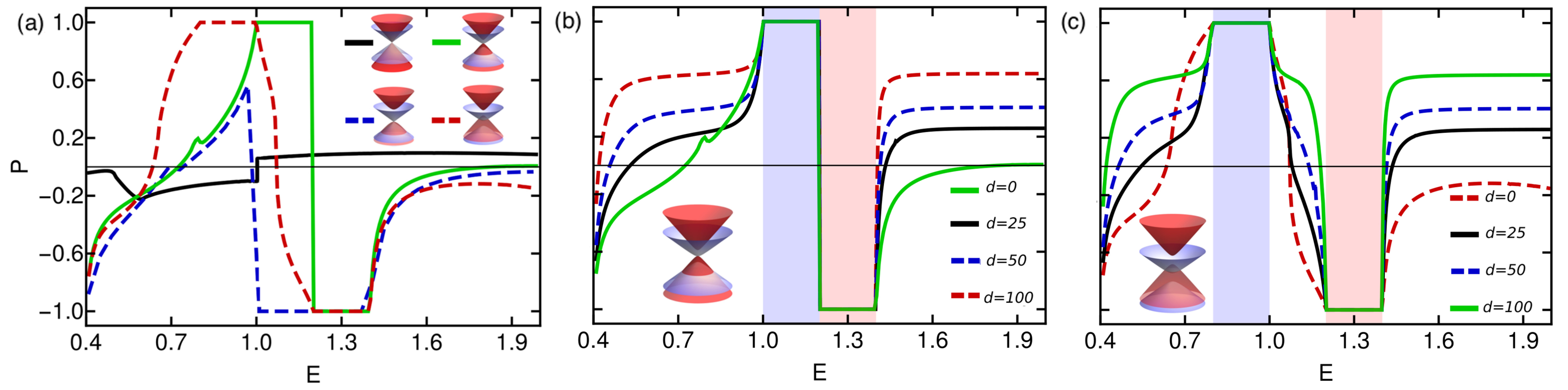}
\caption{Polarization as a function of the Fermi energy $E$ for the cases of massless Dirac fermions in both valleys, chiral symmetry breaking in a single valley, quadratic band crossing point, and valley-orbit coupling in (a). The effect of the smoothness of the junction for the cases of a quadratic band crossing point and valley-orbit coupling in (b) and (c), respectively.}
\label{Pol}
\end{figure*}

We analyze the tunneling as a function of the electron energy $E$ under normal incidence for situations with chiral symmetry breaking, as shown in Fig. \ref{PVFE}(b) to (f). We set the weights $|a^\nu|^2 = 0.5$ in the injected state. Despite a gap opening in the valley $K^+$, the distinctive signature of massless Dirac fermions persists due to the absence of backscattering in the valley $K^-$ for all energies, see Fig.~\ref{PVFE}(b). However, the fraction of massless Dirac fermions in the valley $K^-$ crossing the interface is less than the initial quantity $|a^-|^2=0.5$. As expected, the reflection probability of massive Dirac fermions is nonzero, but the sum $R^++T^+>0.5$. Part of the current flow of massless Dirac fermions converts to massive particles. The fraction of electrons realizing the valley flip and obtaining a non-zero mass is equal to $v_\tau/v_\sigma = 0.1$ independently of the Fermi level. By increasing the energy, massive Dirac fermions are filtered totally in the intraband transmission regime due to the absence of the electronic band for the valley $K^+$. In this energy range, the junction works as a perfect valley filter, see the red shaded region. In the pseudo-ultra-relativistic limit (i.e. at high energies), the reflection and transmission probabilities have an identical behavior compared with the previously discussed gapless case. The reflection and transmissions probabilities are constant for normal incidence and independent of the energy due to the conservation of the pseudospin.

The valley-flip process persists even if the chiral symmetry is broken in both valleys, as shown in Fig.~\ref{PVFE}(c) to (f). The transmission probabilities for the two valleys have the values of 0.6 and 0.4 in the pseudo-ultra-relativistic limit, because the valley-flip process depends only on the Fermi velocities $v_\sigma + \nu v_\tau$ but not on the self-gating potentials $\xi_\nu$ and rest energies $\mu_\nu$. However, the tuning of these parameters can be used to design junctions acting as a valley switch. Depending on the Fermi energy, electrons from a certain valley are filtered, see the blue and red shaded regions in Fig.~\ref{PVFE}(c)-(f). For some topological phase these regions can be separated by a transport gap, see the green shaded region in Fig.~\ref{PVFE}(e) and (f).

To quantify the degree of polarization $P$ in KekGr $pn$ junctions, we define
\begin{equation}
    P = \frac{\langle T^+ \rangle - \langle T^- \rangle }{\langle T^+ \rangle + \langle T^- \rangle },
\end{equation}

\noindent where $\langle T^\nu \rangle$ is the angular averaged transmission of Eq.~\eqref{Tnu}. Figure~\ref{Pol}(a) shows the polarization in junctions for the topological phases of Fig.~\ref{DCs}(a), (c), (e), and (f). We can see that the partial polarization of massless Dirac fermions (black curve) remains almost constant for the intraband transmission ($E > V$) and has the approximated value of $P \approx v_\tau/v_\sigma=0.1$. Note that this asymptotic value of $P$ is unaffected by the smoothness of the junction, being a signature of the valley-flip process that may be detected experimentally. However, the valley polarization of massless Dirac fermions is rather low. A fully valley polarized current flow is achieved by breaking the chiral symmetry. Moreover, the sign of the polarization can be changed by tuning the Fermi energy within the different gaps of the valleys, making these systems a perfect valley switch. Note that the transition is abrupt in the case of a quadratic band crossing point (green curve), but smooth in the case of valley orbit coupling (red dashed curve). 

To investigate the robustness of our findings, we consider a smooth electrostatic potential changing linearly over the distance $2d$ \cite{Allain2011, BetancurOcampo2017a}. In this case the electrons have to tunnel though a forbidden region, which reduces the transmission probability exponentially. However, the polarization of the current flow is unaffected, as shown in Fig. \ref{Pol}(b) and (c). The asymptotic value of $P$ at higher energies increases due to the collimation effect. Electrons cross preferentially under normal incidence because the forbidden region grows for grazing incidence \cite{Cheianov2006, BetancurOcampo2019, ParedesRocha2021}. However, this goes along with an increase of the reflections reducing the overall efficiency of the device. 

In summary, the particular design of Kekul\'e textures in graphene leads to atypical transport phenomena such as valley-cooperative Klein tunneling and valley polarization of currents. We proposed a generalized Hamiltonian in Eq.~\eqref{HKGS} that embodied multiple topological phases with electronic band structures identical to systems with Rashba spin-orbit interactions. We also reported new topological phases such as a Zeeman effect without magnetic field and crossing of conduction bands through the tuning of the effective parameters of the Hamiltonian \eqref{HKGS}, as shown in Fig. \ref{DCs}. In Kekul\'e graphene $pn$ junctions, the chiral symmetry breaking serves as a mechanism to give mass to Dirac fermions through a valley-flip process, similar to the Higgs boson of elementary particles in the standard model. Moreover, these devices can be used to filter perfectly massive Dirac fermions, as seen in Fig. \ref{PVFE} (a) to (c), and obtain fully valley polarized currents, which is the essential building block for valleytronics. Additionally, we showed that the smoothness of the junction does not affect the total polarization and the signal of valley-cooperative tunneling, making it feasible for experimental realizations.

SGyG acknowledges a scholarship from CONACYT. We acknowledge financial support from CONACYT Project A1-S-13469 and the UNAM-PAPIIT research grant IA103020.

\end{document}